\documentstyle[12pt,epsf]{article}
\newcommand{\be}{\begin{equation}}
\newcommand{\ee}{\end{equation}}
\newcommand{\bea}{\begin{eqnarray}}
\newcommand{\eea}{\end{eqnarray}}

\newcommand{\Seff}{S_{\mbox{eff}}}
\newcommand{\Veff}{V_{\mbox{eff}}}
\newcommand{\dsla}{\partial\hspace{-2.2mm}/}

\newcommand{\Asla}{A\hspace{-2.4mm}/}
\begin{document}
\vspace{-1cm}
\noindent
\begin{flushright}
KANAZAWA-99-11\\
KUCP-0139
\end{flushright}
\vspace{20mm}
\begin{center}
{\Large \bf 
Wilson Renormalization Group Equations
\vspace{2mm}\\
for the Critical Dynamics of Chiral Symmetry
}
\vspace*{15mm}\\
\renewcommand{\thefootnote}{\alph{footnote}}
Ken-Ichi Aoki
, Keiichi Morikawa
, Jun-Ichi Sumi$^\dagger$
, Haruhiko Terao 
\\
and Masashi Tomoyose
\vspace*{3mm}\\
Institute for Theoretical Physics, Kanazawa University\\
Kanazawa 920--1192, Japan
\vspace*{3mm}\\
$^\dagger$Department of Fundamental Sciences,\\
Faculty of Ingegrated Human Studies, Kyoto University,\\
Kyoto 606-8501, Japan
\vspace*{40mm}\\
{\large \bf Abstract}
\end{center}

The critical dynamics of the chiral symmetry breaking induced by
gauge interaction is examined in the Wilson renormalization
group framework in comparison with the Schwinger-Dyson approach.
We derive the beta functions for the four-fermi couplings in the
sharp cutoff renormalzation group scheme, from which the critical
couplings and the anomalous dimensions of the fermion composite
operators near criticality are 
immediately obtained. It is also shown that the beta functions lead
to the same critical behavior found by solving the 
so-called ladder Schwinger-Dyson equation, if we restrict the 
radiative corrections to a certain limited type. 
\setcounter{footnote}{0}
\newpage
\pagestyle{plain}
\pagenumbering{arabic}
\noindent
{\large \bf 1. Introduction}
\vspace*{2mm}

The chiral symmetry breaking phenomena has been one of the key issues
to be understood in the non-perturbative dynamics of gauge theories.
The analytical study of this problem has been initiated by the
Nambu-Jona-Lasinio (NJL) model \cite{njl}, 
which was introduced as the effective
theory with four-fermi interactions. For gauge theories particularly the
Schwinger-Dyson (SD) equaitons in the ladder approximation with Landau 
gauge \cite{SDmain,SDfourfermi} have been intensitively studied and 
applied not only to QCD but also to the various models of dynamical 
electroweak symmetry breaking \cite{SDmodel,topcondensation}. 
In QCD, the ladder SD equation with the running gauge
coupling constant, the improved ladder \cite{SDimproved}, 
was found to give good results even qantitatively \cite{SDQCD}. 
However the ladder SD equations are known to 
suffer from some serious problems, specially the strong gauge
dependence \cite{SDdifficulty} and the 
difficulty to proceed beyond the ladder approximation \cite{SDbeyond}. 

On the other hand the Wilson renormalzation group (RG)
\cite{wilsonkogut} has been 
known to offer the powerful method to analyze critical phenomena
and has been applied to the various dynamical problems
mainly in the statistical mechanics. The so-called exact RG equations
\cite{wilsonkogut,wegnerhoughton,exactRG},
which are the concrete formulation of the Wilson RG in the momentum
space, has been recently applied to numerical study of
non-perturbative dynamics in field theories. 
The application to the QCD dynamics has been also considered in this 
framework \cite{quarkmeson}. 
The advantageous features of this method, compared with the SD
approach, are that the critical behavior is analyzed directly
form the RG eqautions, and that it admits the systematic improvement 
of approximation by the derivative expansion and truncation of the
Wilsonian effective action \cite{comoving,souma}. 
Interestingly it is rather recent
that the RG method has been applied to the fermi liquid theory of
superconductvity \cite{fermiliquid}, 
which the NJL model was considered in analogy with. 
Also it should be noted that the fermi liquid theory of high density 
QCD was studied by the RG analyses \cite{denseqcd}.

In this paper we examine the chiral critical dynamics in gauge
theories by using the exact RG equations, specially putting
our attention to the comparison with the SD approach. There have been
known several formulations of the exact RG. Here we simply employ the 
Wegner-Houghton RG equations \cite{wegnerhoughton}, 
which are derived with sharp momentum cutoff,
in the so-called local potential approximation for our present purpose.
The analyses with the exact RG equations with smooth cutoff may be
performed as well \cite{ccbqed}. 
It will be found that the critical behavior is
determined from the beta-functions of the effective four-fermi
couplings induced by gauge interaction with remakably simple
calculation. The phase boundary and also the anomalous dimensions of
the composite operators of fermions near the criticality will be 
evaluated.
Our approximation scheme adopted here is even better than the ladder 
approximation performed in the SD equations on the critical behavior.
Actually, as is seen later, if we make further approximation so as to
pick up only a few types of the radiative corrections, then the critical
behavior is reduced to be identical to that obtained by solving
the ladder SD equation.

\vspace*{4mm}
\noindent
{\large \bf 2. Scheme of the RG equations}
\vspace*{2mm}

The ladder SD equation for the fermion mass function is given in the 
form of an integral equation, where the momentum integration is
carried out with sharp cutoff. In order to see the direct relation 
between the critical dynamics obtained by the two methods; the SD 
equation and the RG equation, we consider the exact RG equation
defined with sharp cutoff in this paper. There have been known the several 
formalisms for the exact RG \cite{wilsonkogut,exactRG}. 
Here we shall adopt the so-called Wegner-Houghton
RGE \cite{wegnerhoughton} derived as follows.

If we devide the freedom of the quantum field $\phi(p)$ into the
higher freqency modes with $|p|>\Lambda$ and the lower frequency modes 
with $|p|<\Lambda$ by introducing the cutoff scale $\Lambda$ in the 
Euclidean formalism,
then the Wilsonian effective action at this scale, $\Seff[\phi;\Lambda]$, 
may be defined by integrating out the higher frequency modes in the 
partition function. Namely
\be
Z=
\int \prod_{|p|<\Lambda_0}d\phi(p)~e^{-S_0[\phi;\Lambda_0]}
=\int \prod_{|p|<\Lambda}d\phi(p)~e^{-S_{\mbox{\tiny eff}}[\phi;\Lambda]},
\ee
where $S_0$ denotes the bare action with the bare cutoff $\Lambda_0$.
This effective action contains the general operators invariant
under the original symmetries in the bare action, for example the
chiral symmetry of our present concern. 

The Wegner-Houghton RGE determines the variation of the Wilsonian
effctive action under the infinitesimal change of the cutoff $\Lambda$.
For example, the RGE for the $D$-dimensional scalar field theory
is found to be given exactly as
\bea
& &\frac{\partial \Seff}{\partial t}
=D\Seff-
\int \frac{d^Dp}{(2\pi)^D}~\phi_p\left(
\frac{2-D-\eta}{2}-p^{\mu}\frac{\partial'}{\partial p^{\mu}}
\right)
\frac{\delta\Seff}{\delta \phi_p} \nonumber \\
& &\hspace*{14mm}
-\frac{1}{2}\int \frac{d^Dp}{(2\pi)^D}\delta(|p|-1) 
\left\{
\frac{\delta \Seff}{\delta\phi_p}
\left(
\frac{\delta^2 \Seff}{\delta\phi_p\delta\phi_{-p}}
\right)^{-1}
\frac{\delta \Seff}{\delta\phi_{-p}}
-\mbox{\rm tr}\ln 
\left(
\frac{\delta^2 \Seff}{\delta\phi_p\delta\phi_{-p}}
\right)
\right\},
\eea
where $t=\ln(\Lambda_0/\Lambda)$ is introduced as the scale parameter.
The 1st line of the RGE represents nothing but the canonical scaling 
of the effective
action. While the 2nd line comes from the radiative corrections which
correspond to the tree and the 1-loop Feynman diagrams including only 
the propagators with the momentum of the scale $\Lambda$.

In the practical analysis it is inevitable to simplify this RGE by
some approximation. Here we shall examine the RGE in the so-called 
local potential approximation \cite{wegnerhoughton,lpa}.
In this approximation the radiative corrections to any operators containing
derivatives are ignored in the RGE (2). Therefore solely the potential part
of $\Seff$, $\Veff$, may be evolved with the shift of $\Lambda$. It should be
noted that the wave function renormalization is ignored in this scheme.
The RGE for the scalar theory is given explicitly by
\be 
\frac{\partial \Veff}{\partial t}=D\Veff-\frac{D-2}{2}\phi
\frac{\partial \Veff}{\partial \phi}
+ \frac{A_D}{2}\ln \left(1+\frac{\partial^2 \Veff}{\partial \phi^2}\right),
\ee
where $A_D=2/(4\pi)^{D/2}\Gamma(D/2)$ is the factor from the momentum
integration. This equation offers us a set of the infinitely many beta
functions for the general couplings appearing in $\Veff$. However it should
be noted here that each beta function may be evaluated through just one loop
corrections with the general effective interactions. The non-perturbative
nature of the RGE is supposed to be maintained by solving the infinitely
many coupled renormalization equations. Actually it has been known that
this approximated RGE is quite effective in the case of the scalar theories
\cite{comoving,souma}.
The generalization of this RGE to include fermions has been also
studied in the relation with the triviality-stability bound for the
Higgs boson mass in the standard model
\cite{yukawaRG}.

Now let us consider to apply this formulation to the massless fermions
coupled by gauge interaction. For example we may take the action of the
massless QED as the bare action
\be
S_0=\int d^4 x \left\{
\bar{\psi}\dsla \psi
+e\bar{\psi}\Asla \psi+\frac{1}{4}F_{\mu\nu}^2
+\frac{1}{2\alpha}(\partial_{\mu}A_{\mu})^2
\right\},
\ee
where $\alpha$ is the gauge parameter.
As is well known, the gauge invariance is not maintained anymore, once
the momentum cutoff is performed. Therefore the generic gauge
non-invariant operators are generated in the Wilsonian effective action.
Then we must encounter the rather complicated problem to pick up the
special RG flows corresponding to the gauge invariant theories in the
infinite dimensional coupling space. Recently it has been discussed
how to deal with the gauge theories in the framework of the Wilson
RG by using the modified Slavnov-Taylor
identities \cite{MSTI}. 

Here, however, we shall simply ignore the corrections to the operators 
including the gauge fields as well as imposing the local potential 
approximation as the first step towards the analysis of the chiral 
symmetry breaking phenomena.
Then we may avoid the intriguing problem of the gauge invariance, since
any gauge non-invariant operators do not appear in the effective action.
This approximation is indeed so rough as to make the beta function of the
gauge coupling vanish identically, therefore it cannot be supposed to
give any picture of the real dynamics of the gauge theories. However,
on the other hand, the so-called ladder approximaton used in the 
SD approach also totally ignores the vertex corrections
as well as the corrections to the gauge kinetic functions. Therefore it
would be meaningful to examine the RGE in this scheme in comparison with
the SDE's in the ladder approximation. The effect of the running gauge
coupling will be discussed in section 4.
Nevertheless this approximation scheme is thus rather crude, it will
be seen that the chiral critical behavior may well be described. 
Actually it will be found that this approximation is even better than the 
ladder approximation applied for the SD equations. 

The Wilsonian effctive action to be solved by the RGE in this scheme 
is now reduced to the form of
\be
\Seff[\psi,\bar{\psi};\Lambda]=\int d^4 x \left\{
\bar{\psi}\dsla \psi+\Veff(\psi,\bar{\psi};\Lambda)
+e\bar{\psi}\Asla \psi+\frac{1}{4}F_{\mu\nu}^2 
+\frac{1}{2\alpha}(\partial_{\mu}A_{\mu})^2
\right\},
\ee
where $\Veff(\psi, \bar{\psi})$ denotes the general potential composed of
the chiral symmetric multi-fermion operators. These multi-fermion
operators, which are induced by exchange of the ``photon'' with higher 
momentum. 
The so-called gauged NJL model is often examined in the SD approach
and the phase diagram in the two parameter
space of the gauge coupling and the four-fermi coupling has been
examined \cite{SDfourfermi}. 
However, in the RG point of view, this coupling space should 
be regarded as a subspace of the infinite dimensional coupling space
of the Wilsonian effective action.  
It should be noted also that these multi-fermi operators are
irrelevant or non-renormalizable, and therefore, are not considered 
in the perturbative QED. However they cannot be simply 
discarded in the strong coupling region. It will be seen in the next
section that the four-fermi coupling turns out to be relevant near
the criticality and plays a crutial role for the critical 
dynamics of the chiral symmetry breaking. 

\vspace*{4mm}
\noindent
{\large \bf 3. Critical dynamics of the chiral symmetry}
\vspace*{2mm}

In this section we examine explicitly the Wegner-Houghton RGE in the
approximation discussed in the previous section. The form of the effective
potential $\Veff$ written in terms of the fermions is found to be
restricted into a polynomial composed of the following parity and 
chiral invariant operators, which are mutually independent;
\bea
& &~~{\cal O}_1 =
(\bar{\psi}\psi)^2+(\bar{\psi}i\gamma_{5}\psi)^2
= -\frac{1}{2}\left\{
(\bar{\psi}\gamma_{\mu}\psi)^2-(\bar{\psi}\gamma_{5}\gamma_{\mu}\psi)^2
\right\}, \nonumber \\
& &~~{\cal O}_2 =
(\bar{\psi}\gamma_{\mu}\psi)^2+(\bar{\psi}\gamma_{5}\gamma_{\mu}\psi)^2,\\
& &~~{\cal O}_3 =
\left\{
(\bar{\psi}\gamma_{\mu}\psi)(\bar{\psi}\gamma_{5}\gamma_{\mu}\psi)
\right\}^2. \nonumber
\eea
Therefore the 4-fermi part of the effective potential may be written down as
\be
\Veff(\psi,\bar{\psi};t)=
-\frac{G_S(t)}{2\Lambda^2}\left\{
(\bar{\psi}\psi)^2+(\bar{\psi}i\gamma_{5}\psi)^2
\right\}
+\frac{G_V(t)}{2\Lambda^2}\left\{
(\bar{\psi}\gamma_{\mu}\psi)^2+(\bar{\psi}\gamma_{5}\gamma_{\mu}\psi)^2
\right\}.
\ee
Hereafter let us call $G_S$ the scalar four-fermi coupling and also $G_V$
the vector four-fermi coupling.
\footnote{The sign of the scalar four-fermi coupling introduced here 
follows the conventional one in the literatures.}

Now let us evaluate the radiative corrections to these four-fermi operators,
since it will be found enough to see only these couplings for the
purpose of understanding the critical dynamics. By using the propagator
of the gauge field in the Landau gauge $(\alpha=0)$, the RG equations 
for the four-fermi couplings are found to be
\bea
& &\frac{d}{dt}g_S=-2g_S+\frac{3}{2}g_S^2+4g_Sg_V
+g_S\lambda -\frac{1}{6}\lambda^2, \nonumber \\
& &\frac{d}{dt}g_V= -2g_V+\frac{1}{4}g_S^2
-g_V\lambda -\frac{1}{12}\lambda^2,
\eea
where we introduced the rescaled couplings,
$g_S=G_S/(4\pi^2)$,  $g_V=G_V/(4\pi^2)$, $\lambda=3e^2/(4\pi^2)$.  
Here we should note that any multi-fermi couplings more than four do not
take part in the radiative corrections for the four-fermi couplings owing
to the 1-loop nature of the RGE. Therefore we may obtain the RG flows
within the 2 dimensional coupling space (or 3 dimensional, if
the gauge coupling is also taken into account) irrespectively to other
couplings. 

In Fig.1 the Feynman diagrams representing the one-loop corrections to
the four fermi couplings are shown. 
Let us call the corrections
given by the diagrams in the dashed-line box in Fig.1
the ``ladder'' type, and the others the ``non-ladder'' type
hereafter. 
If we approximate the RGE by restricting to the ``ladder
type'' correction, then the beta function for the scalar four-fermi coupling 
is found to be given by 
\be
\frac{d}{dt}g_S=-2g_S+2(g_S + \lambda/4)^2,
\ee
where it is noted that the RGE for the scalar four-fermi decouples from that
for the vector four-fermi coupling.

\begin{figure}[hbt]
\epsfxsize=0.8\textwidth
\begin{center}
\leavevmode
\epsffile{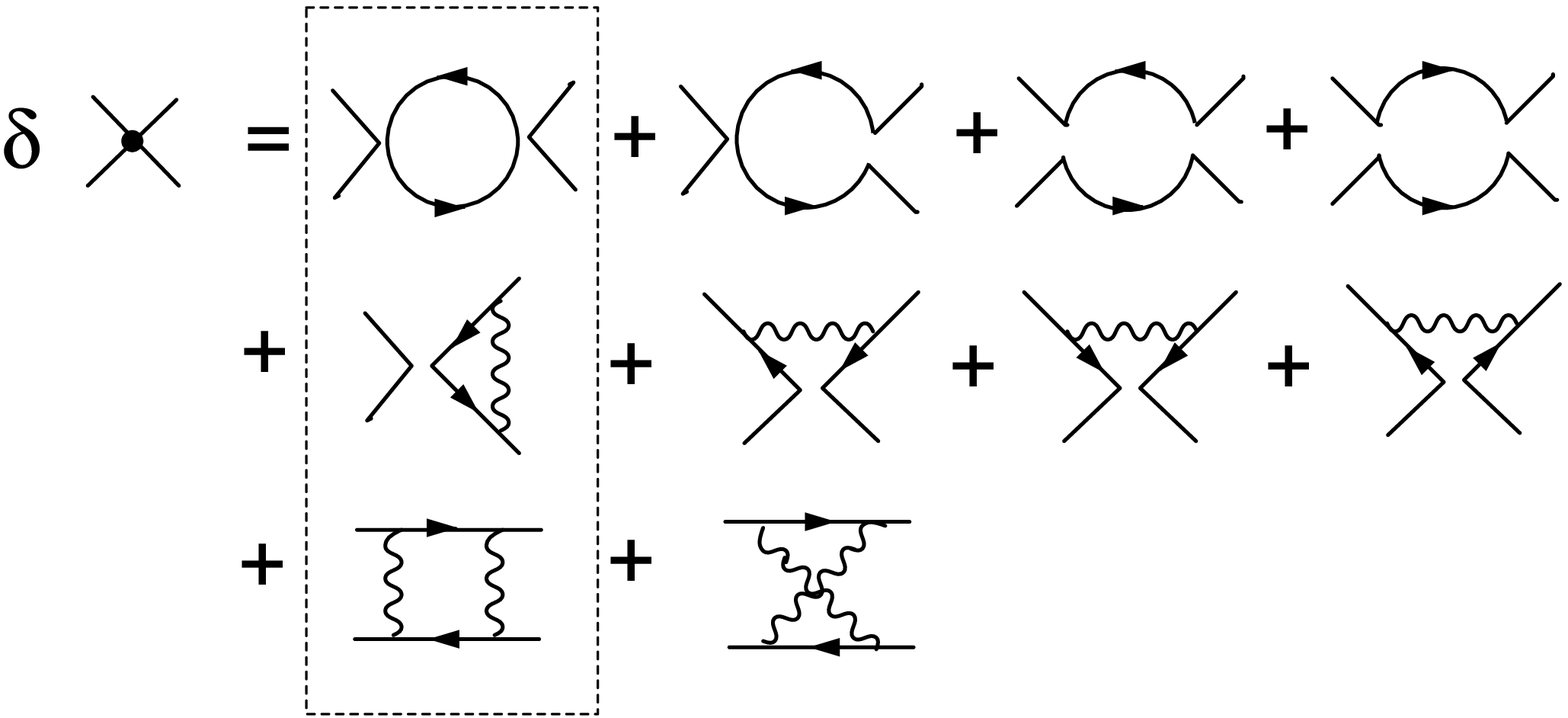}
\vspace{3mm}\\
\parbox{130mm}{
Fig.1: Feynman diagrams of the radiative corrections to the four-fermi
couplings $g_S$ and $g_V$ considered in the RGE (8). The diagrams 
surrounded by the dashed line show the ``ladder'' type corrections.
}
\end{center}
\end{figure}

Before examining the full RGE's (8), let us consider the RG flows
in the subspace of $\lambda=0$.
The beta functions tell the ``fixed points'' at 
$(g_S^*, g_V^*)=(0,0), (1, 1/8), (-4, 2)$.
$(0,0)$ is the IR trivial fixed point, and $(1, 1/8)$ is the UV
fixed point on the critical surface.
\footnote{
Strictly speaking $(g_S^*, g_V^*)=(1, 1/8)$ is not a fixed point, 
since the beta function for the eight-fermi coupling does not vanish 
there. However it turns out to be the non-trivial fixed point for the 
space-time dimension of $2<d<4$. 
We do not consider the point $(-4, 2)$, since it seems to be fake due 
to this approximation.
}
The RG flows are found as is shown in Fig.2. It is seen that there are
two phases divided by a critical surface. The chiral symmetry is supposed 
to be spontaneously broken in the upper region in Fig.2. Then we may 
realize that the chiral symmetry breaking is caused essentially by the 
strong scalar four-fermi interaction, not by the vector four-fermi 
interaction.

It is also easy to evaluate the exponents which are important physical 
quantities in the critical dynamics. By linearlizing the RG equations 
around the UV ``fixed point'', the dimensions of the relevant 
four-fermi coupling and the irrelevant four-fermi coupling are found 
to be $2$ and $-5/2$ respectively.
The relevant four-fermi operator is given by the combination of
${\cal O}_{\mbox{rel}}={\cal O}_1-(1/8){\cal O}_2$. The renormalized
trajectory in Fig.2 is given by the straight line passing through the 
non-trivial fixed point. Indeed we may deduce from the RGE's (8)
\be
\frac{d}{dt}(g_S-8g_V)=-(2+\frac{1}{2}g_S)(g_S-8g_V),
\ee
which means that once the combination of $(g_S-8g_V)$ is vanishing at a 
point, then it keeps null along the renormalization flow. Therefore the 
renormalized trajectory is precisely given by the line of $g_S=8g_V$. 
Namely the effective four-fermi operator in the low energy is just 
${\cal O}_{\mbox{rel}}$ irrespectively to the phases. 

Next let us examine the RG equations (9) given by the ``ladder type'' 
corrections.
For each gauge coupling $\lambda$ there are the UV fixed point and the
IR fixed point, which are found to be
\be
g_S^*(\lambda)=\left(1 \pm \sqrt{1-\lambda}\right)^2/4,
\ee
where $+$ is for the UV fixed point and $-$ is for the IR fixed point.
Namely they form a fixed line in the $(\lambda, g_S)$ space as 
shown in Fig.3. The phase boundary is also shown in Fig.3. 
The upper region is supposed to be the chiral symmetry broken phase. 
The critical gauge coupling is given by
\be
\lambda_{\mbox{cr}}=1.
\ee 
Indeed this critical surface just coinsides with that 
obtained by solving the SDE in the ladder approximation
\cite{SDmain,SDfourfermi}.

The anomalous dimension of the four-fermi coupling $g_S$, 
$\gamma_G=2+\mbox{dim}[g_S]$, near criticality is immediately
deduced from the RGE (9) as
\be
\gamma_G=4g_S^*(\lambda)+\lambda
=2\left(1 + \sqrt{1 - \lambda}\right),
\ee
which is also found to coincide with the result
by the ladder SDE \cite{SDfourfermi}. 
Therefore it is seen that our approximation used to derive the RGE's (8) 
is certainly better than the ladder approximation.
Moreover it is easy to take the all corrections shown by Fig.1 including
the ``ladder type'' ones in our framework.
It should be noted that the sum of corrections of the ``ladder'' diagram
and the ``crossed ladder'' given in the last line of Fig.1 is found to
be free from gauge parameter dependence. Thus this extention of approximation
beyond ``ladder'' is significant to obtain the gauge independent
results \cite{ccbqed}. 
We would like to stress here that the exact RG equations allow us to
examine the critical dynamics by remarkably simple calculation, 
which is a clear contrast with the SD approach.

\begin{figure}[hbt]
\epsfxsize=0.5\textwidth
\leavevmode
\epsffile{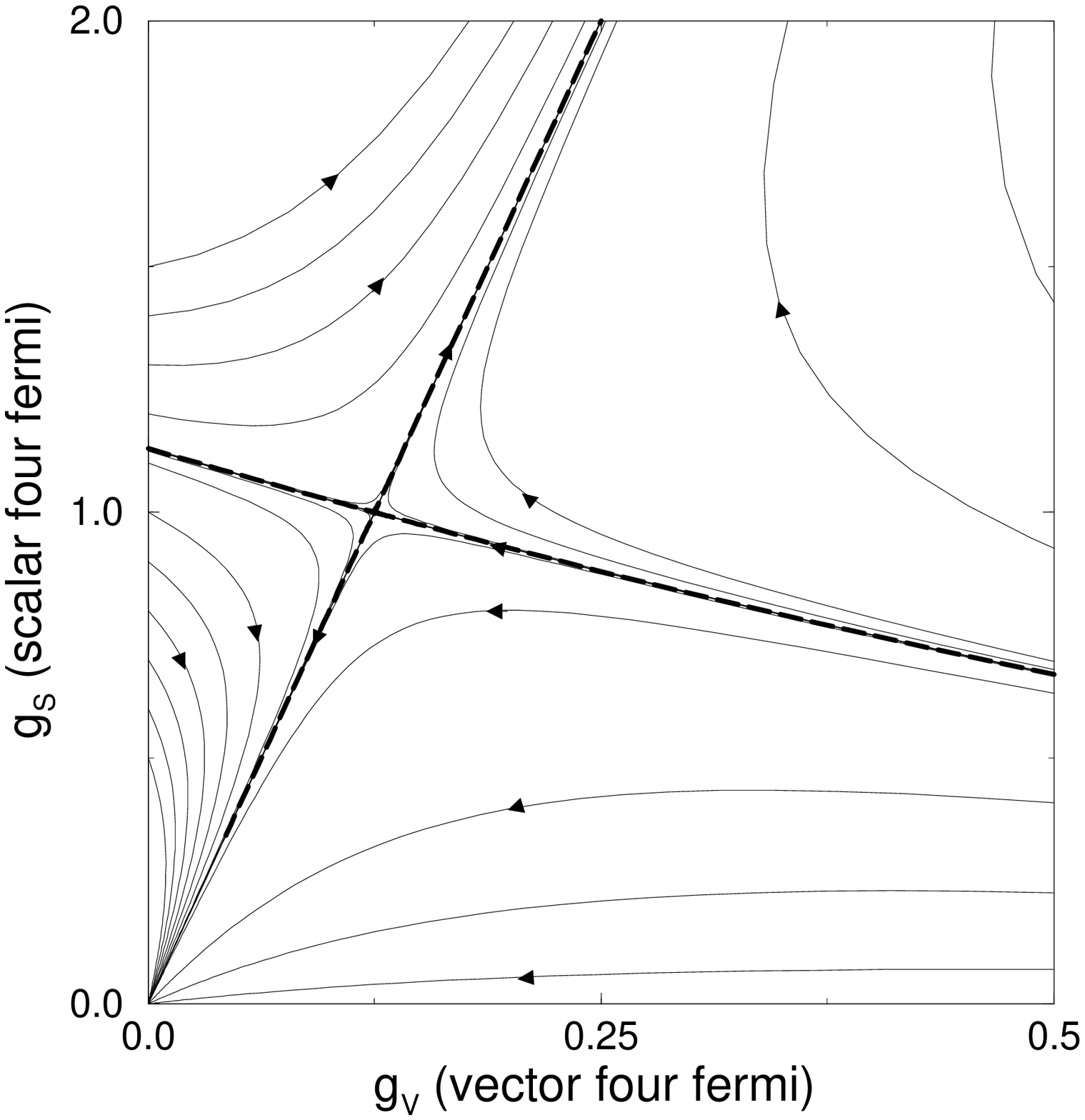}
\epsfxsize=0.5\textwidth
\leavevmode
\epsffile{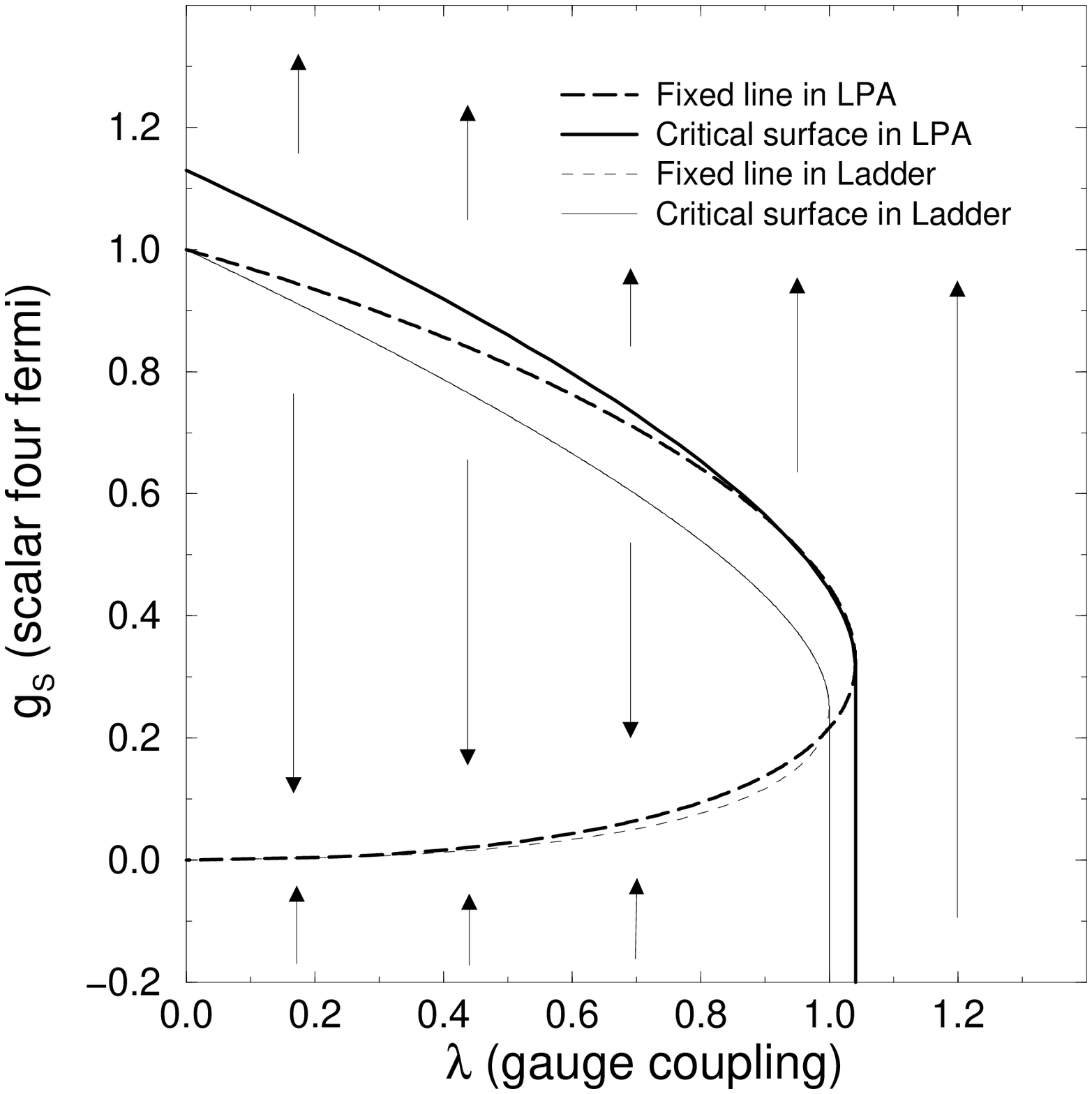}
\begin{center}
\parbox[t]{75mm}{
Fig.2: RG flow diagram of the four-fermi couplings $(g_V, g_S)$ in 
the $e=0$ plane. The critical surface and the renormalized trajectory
are shown by dashed lines.
}
\hspace*{5mm}
\parbox[t]{75mm}{
Fig.3: The fixed lines projected on the $(\lambda, g_S)$-plane and the 
critical couplings of the scalar four-fermi obtained from the beta 
functions given in eq.(8) and in eq.(9).
}
\end{center}
\end{figure}

Now we are in a position to go beyond ``ladder'' by examining the full
RG equations (8). The fixed line $(g_S^*(\lambda),g_V^*(\lambda))$, 
which is given by the solution of the 3rd order equation in turn, 
is shown in Fig.3 and also in Fig.4 by projection to 
the $(\lambda, g_S)$-plane and $(g_V, g_S)$-plane respectively. 
It is seen that the critical gauge coupling
is now found to be slightly bigger than the value in the ``ladder'' 
approximation, ($\lambda_{\mbox{cr}}=1.0409$).
In Fig.4 the critical surface separating the two phases is also shown by 
the cross sections at various gauge coupling up to the critical one.
The critical surface given in the case of ``ladder'' type should be 
compared with the cross section between the critical surface and the 
$g_V=0$ plane, which is found as shown in Fig.3. 
It is seen that the phase boundary obtained by our scheme is 
shifted towards outside compared with the phase boundary, which has
been known so far in the ladder SD approach.

The exponents at the fixed line also are similarly obtained. The
exponent or the dimension
of the relevant operator, which was found to be $2$ in the previous 
analysis for $\lambda=0$, reduces as the gauge coupling becomes larger.
Then it eventually vanishes at $\lambda=\lambda_{\mbox{cr}}$, which is also 
seen directly from the eq.(8). In Fig.5 the anomalous dimension of the
relevant four-fermi coupling $\gamma_G$ is presented in comparison with 
the ``ladder'' value given by eq.(14).

Before ending this section let us mention the anomalous dimension of the 
fermion mass operator, which we denote $\gamma_m$. In order to evaluate it we 
may incorporate the mass term in the effective action. Then the beta 
function for the mass $m$ may be derived by one-loop diagrams and is found 
to be
\be
\frac{d}{dt}m= m - \frac{2m}{1+m^2}(g_S + \lambda/4).
\ee
Here it should be noted that the contribution from the ``non-ladder''
type corrections vanishes, therefore the vector four
fermi coupling does not appear in this beta function. 
The anomalous dimension on the fixed
line is simply given by
\be
\gamma_m=2g_S^*(\lambda)+\frac{\lambda}{2},
\ee 
which is shown in Fig.5 as well. In the ``ladder'' case it is seen
that the anomalous dimensions satisfy the relation, 
$\gamma_m=\gamma_G/2$, which has been known also in the analysis of
the ladder SDE \cite{SDfourfermi}. 
In our analysis, however, $\gamma_m$ turns out to be fairly larger than 
$\gamma_G/2$ and also than $\gamma_m$ obtained so far in the ladder SD 
approach.

\begin{figure}[hbt]
\epsfxsize=0.5\textwidth
\leavevmode
\epsffile{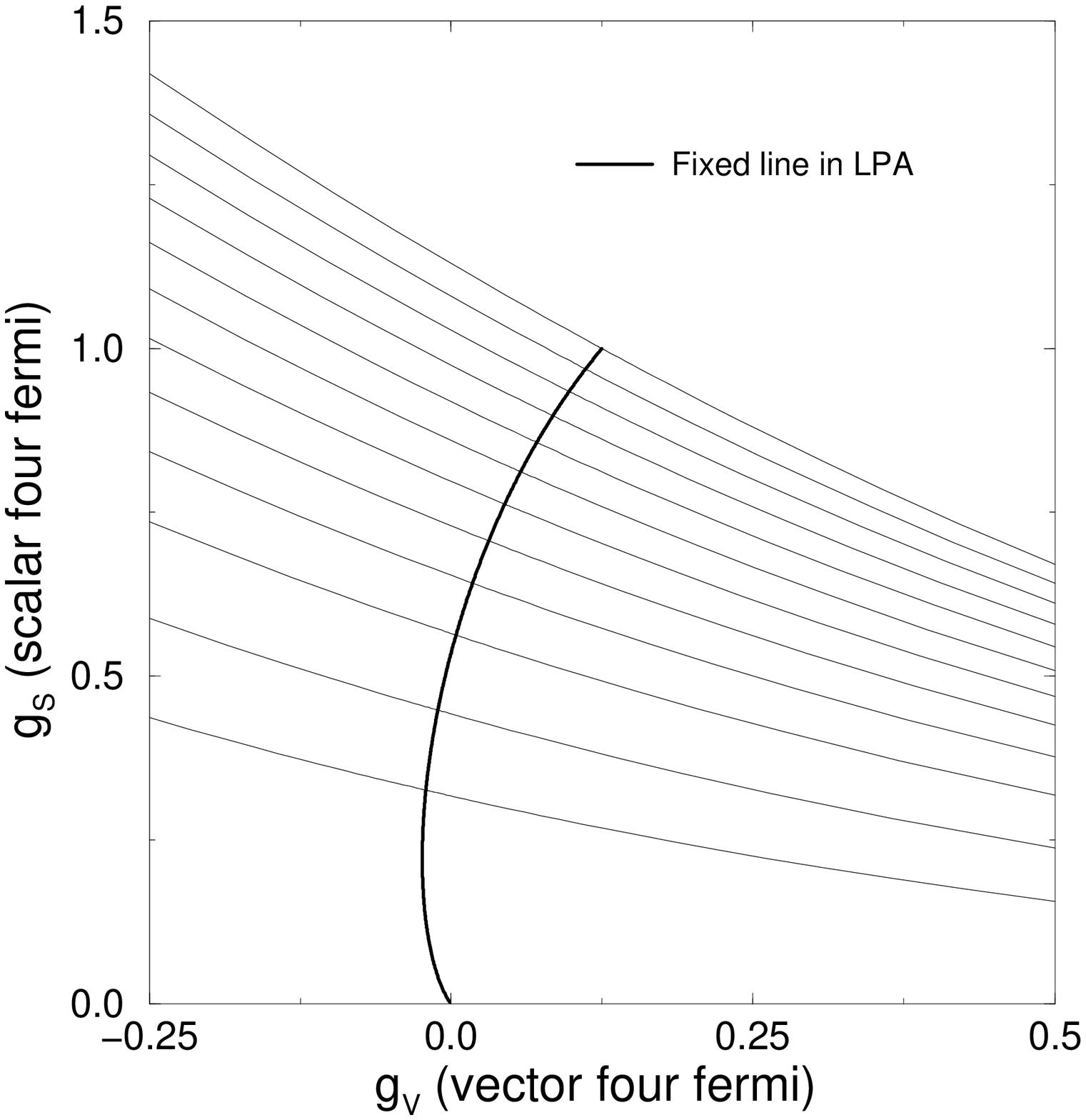}
\epsfxsize=0.5\textwidth
\leavevmode
\epsffile{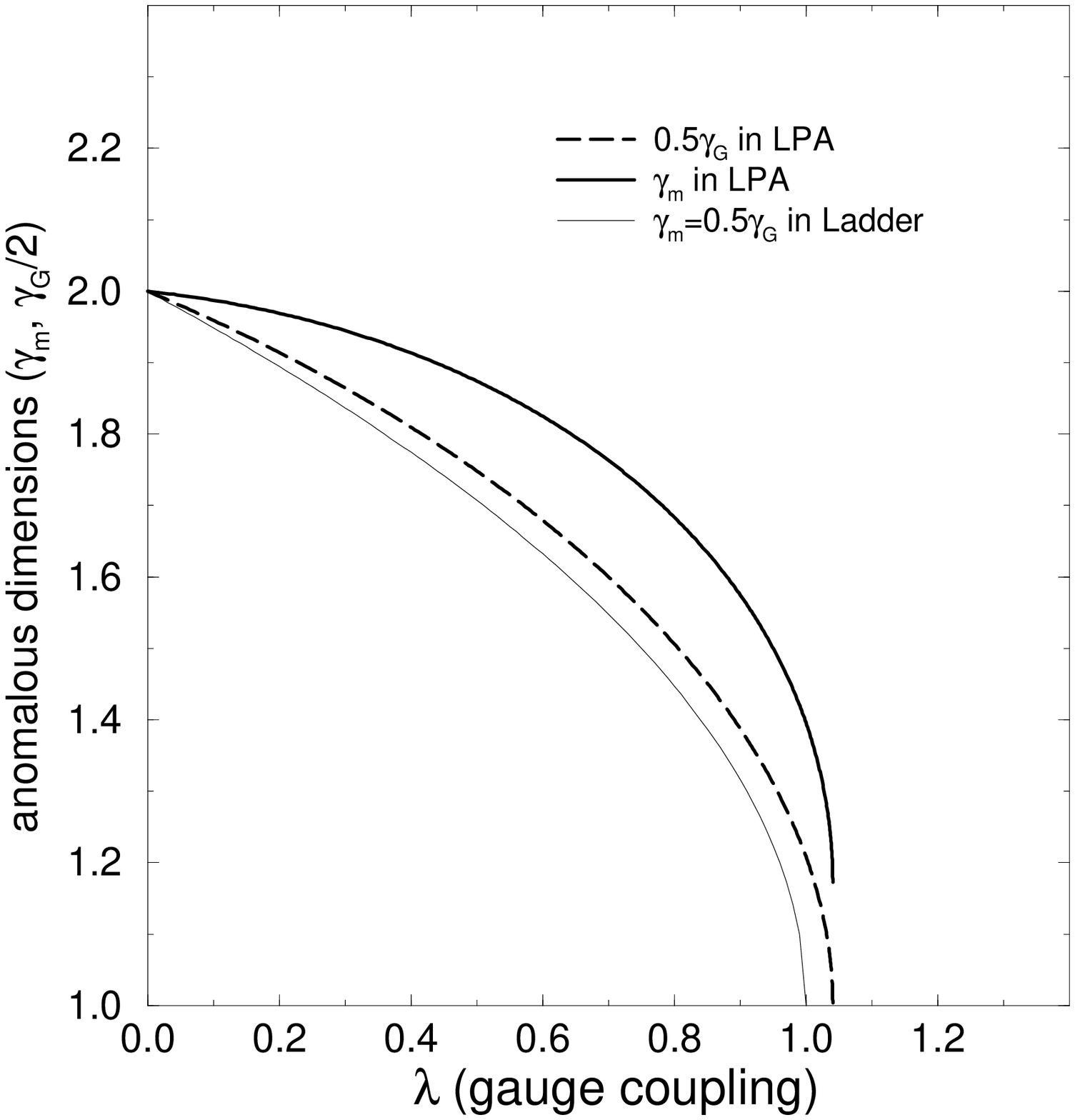}
\begin{center}
\parbox[t]{75mm}{
Fig.4: Cross sections of the critical surface at 
$\lambda=0,0.1,0.2,\cdots,1.0, \lambda_{\mbox{cr}}$
and the fixed line projected on the plane.
}
\hspace*{5mm}
\parbox[t]{75mm}{
Fig.5: Anomalous dimensions $\gamma_m$ and $\gamma_G$
obtained in our LPA scheme and in the ladder approximation.
}
\end{center}
\end{figure}

\vspace*{4mm}
\noindent
{\large \bf 4. RG flow with the running gauge coupling}
\vspace*{2mm}

So far the effect of the renormalization of the gauge coupling has
been totally ignored, therefore the obtained phase diagrams do not 
reflect the realistic ones for gauge theories. 
In the SD approach, the so-called improved ladder approximation
\cite{SDimproved},
in which the gauge coupling is simply replaced by the running coupling 
subject to the perturbative RGE apart from the SD framework, has been
often used. However this prescription cannot be regarded as systematic 
improvement of the approximation. On the other hand, 
in the Wilson RG framework it is possible to include the correction to 
the gauge coupling naturally by improvement of the previous approximation.
This makes a clear contrast to the improved ladder approximation.

If we try to treat the non-perturbative dynamics by strong gauge 
interactions faithfully in the Wilson RG framework, 
we must encounter the
hard problems such as extraction of the gauge invariant theories,
development of simple approximation scheme, incorporation of the 
topological excitations and so on.
However, as a primitive approximation, we may evaluate the Wilson 
beta function of the gauge coupling by the 1st order correction, 
namely the perturbative one. Then it is enough to solve the RG
equations given by (8) in turn coupled with the perturbative RG
equation for the gauge coupling $\lambda$. In 
Fig.6 the RG flows for QED obtained in this manner are shown in the
$(\lambda, g_S)$-plane.
The critical surface separating the spontaneously broken and
the unbroken phases is maintained, while the non-trivial fixed points
turn out disappear. Note that the point 
$(\lambda, g_S^*, g_V^*)=(0, 1, 1/8)$ is not a fixed point in 4 dimensions. 

The RG flows for the QCD like asymptotically free gauge theory is shown 
in Fig.7. 
These results should be compared with those obtained by
solving the SD equations \cite{ksy}.
It is seen that the phase structure is completely swept off. 
The entire reagion is supposed to be in the broken phase of the chiral 
symmetry, since the 4-fermi coupling keeps growing in the infrared. 
The effective theories on the attracting line coming out from
the trivial fixed point correspond to the continuum limit of QCD.
Namely this line gives the so-called renormalized trajectory of QCD. 
However other flows of the gauged NJL models, specially starting at
the critical point of the NJL model in the UV limit, do not 
converge on the renormalized trjectory.
Therefore it may be supposed that these flows show other
renormalized trajectories. If this is the case, the gauged NJL model 
offers non-perturbatively renormalizable theories different form QCD.
Indeed existence of the non-trivial continuum limit other than
QCD, or renormalizability of some kinds of the gauged NJL models
has been claimed so far \cite{ksy,hkkn}.
In our frame work of the non-perturbative RG, renormalizability of the
gauged NJL model may be shown by examining whether these flows are
really the renormalized trajectories or not.    
Such studies will be reported separately \cite{paper3}.

\begin{figure}[hbt]
\epsfxsize=0.5\textwidth
\leavevmode
\epsffile{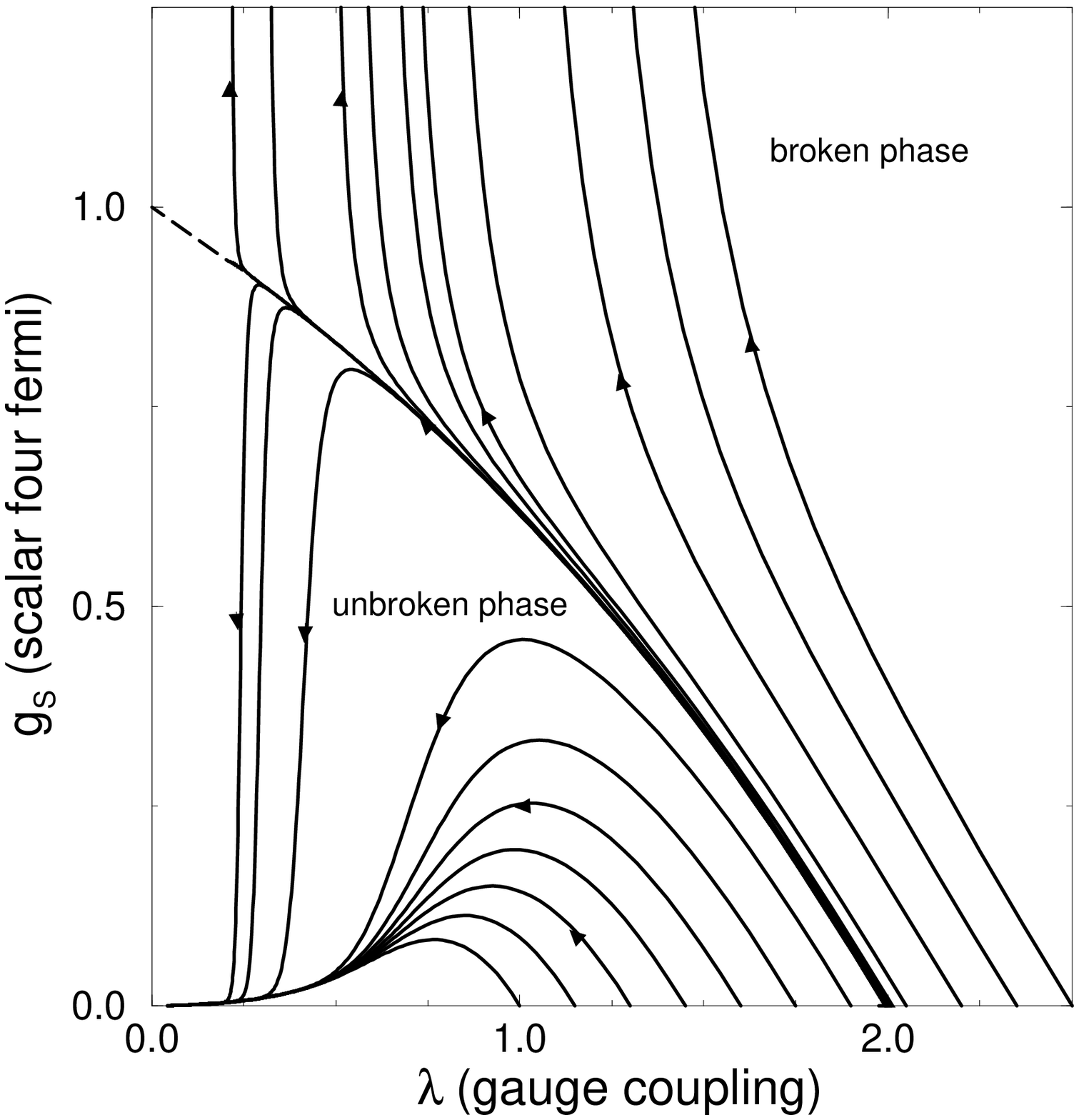}
\epsfxsize=0.5\textwidth
\leavevmode
\epsffile{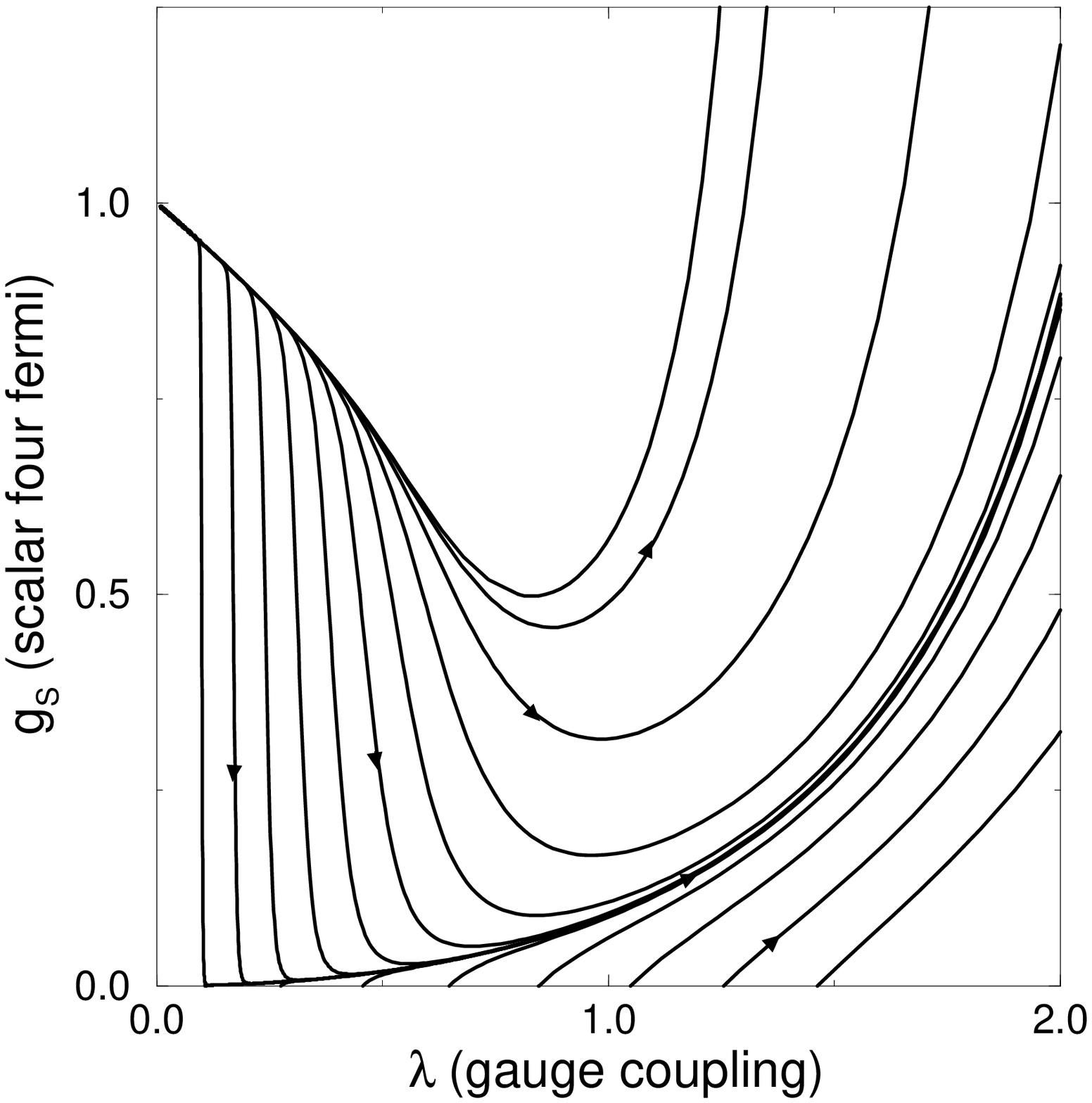}
\begin{center}
\parbox[t]{75mm}{
Fig.6: RG flow diagram for QED. There appears the two phase structure.
The upper (lower) region is supposed to be (un)broken phase.
}
\hspace*{5mm}
\parbox[t]{75mm}{
Fig.7. RG flow diagram for the QCD like gauge theory.
There appears no phase boundary.
}
\end{center}
\end{figure}

\vspace{4mm}
\noindent
{\large \bf 5. Discussions}
\vspace*{2mm}

In this paper we examined the chiral critical behavior of the gauge
theories in the Wilson RG framework. We considered evolution of 
the effective potential composed of the chirally invariant multi-fermi 
operators by the exact RG equation with sharp cutoff in the local 
potential approximation. The RG flow of the four-fermi couplings
were found to determine the phase structure.  
It is straightforwards and remarkably easy to find the critical
surface and the anomalous dimensions of the composite operators of
fermions in this framework. Moreover the critical dynamics obtained by 
solving the ladder SD equations are exactly reproduced by restricting
the radiative corrections taken in the beta functions to the
``ladder'' type. While our RG equations contain also the
``non-ladder'' type corrections, which are neccessary to obtain 
gauge independent physical results \cite{ccbqed}, though 
we have considered only the case of the Landau gauge in this paper. 

However we cannot assert with this analysis that the chiral symmetry 
is indeed spontaneously broken in the region supposed to be the broken 
phase. In order to clarify it we need to evaluate the order parameters 
such as the dynamical mass of the fermion, the condensation of the 
composite operator of fermion, and so on. Evaluation of these order
parameters is important especially in QCD, because they are 
the physical quantities to show the dynamical chiral symmetry breaking. 
While, for example, the spontaneous generation of the fermion mass 
itself seems to be even non-trivial in the Wilson RG picture, since 
the Wilsonian effective action remains chiral symmetric in evolution, 
that is, there is no room for the mass term to show up. 
These issues as well as the method to evaluate the order parameters 
in the Wilson RG framework will be also reported in a separate 
publication \cite{paper2}.

In our analyses the RG eqaution for the gauge coupling was
approximated by the perturbative one.
Of course the fully non-perturbative treatment for the RGE is required
to see the dynamics of strong gauge interaction in infrared.
It would be still an open question whether the Wilson RG approach
gives a useful framework in this non-perturbative region. 
However it may be said that the non-perturbative RG has a good chance 
to seek for the dynamical chiral symmetry breaking phenomena in 
gauge theories further by going beyond the level examined so far 
in the SD approach.




\begin{thebibliography}{99}
\bibitem{njl}
Y.~Nambu and G.~Jona-Lasinio, Phys.~Rev. {\bf 122} (1961) 345.
\bibitem{SDmain}
T.~Maskawa and H.~Nakajima, Prog.~Theor.~Phys. {\bf 52} (1974) 1326.
R.~Fukuda and T.~Kugo, Nucl.~Phys. {\bf B117} (1976) 250.
V.~A.~Miransky, Nuovo Cim. {\bf 90A} (1985) 149.
\bibitem{SDfourfermi}
K.-I.~Kondo, H.~Mino and K.~Yamawaki, Phys.~Rev. {\bf D39} (1989) 2430.
K.~Yamawaki, {\it in Proc. Johns Hopkins Workshop on Current Problems
in Particle Theory 12, Baltimore, 1988},
eds. G.~Domokos and S.~Kovesi-Domokos (World Scientific, Singapore, 1988).
T.~Appelquist, M.~Soldate, T.~Takeuchi and L.C.R.~Wijewardhana, {\it ibid}.
W.~A.~Bardeen, C.~N.~Leung and S.~T.~Love, Phys.~Rev.~Lett.
{\bf 56} (1986) 1230.
C.~N.~Leung, S.~T.~Love and W.~A.~Bardeen, Nucl.~Phys. {\bf B273}
(1986) 649.
\bibitem{SDmodel}
K.~Yamawaki, M.~Bando and K.~Matumoto, Phys.~Rev.~Lett.
{\bf 56} (1986) 1335.
T.~Akiba and T.~Yanagida, Phys.~Lett. {\bf B169} (1986) 432.
B.~Holdom, Phys.~Rev. {\bf D24} (1981) 1441.
\bibitem{topcondensation}
V.~A.~Miransky, M.~Tanabashi and K.~Yamawaki, Phys.~Lett.
{\bf B221} (1989) 177.
W.~A.~Bardeen, C.~T.~Hill and M.~Lindner, Phys.~Rev. {\bf D41} (1990) 1647.
B.~Holdom, Phys.~Rev. {\bf D54} (1996) 1068.
K.~Yamawaki, hep-ph/9603293,
{\it in the proceedings of 14th Symposium on Theoretical Physics:
Dynamical Symmetry Breaking and Effective Field Theory, Cheju, Korea, 21-26 Jul
1995}.
\bibitem{SDimproved}
V.~A.~Miransky,~Sov.~J.~Nucl.~Phys. {\bf 38} (1984) 280.
K.~Higashijima, Phys.~Rev. {\bf D29} (1984) 1228.
\bibitem{SDQCD}
K-I.~Aoki, T.~Kugo and M.~G.~ Mitchard, Phys.~Lett. {\bf B266} (1991) 467.
K-I.~Aoki, M.~Bando, T.~Kugo, M.~G.~Mitchard and H.~Nakatani,
Prog.~Theor.~Phys. {\bf 84} (1990) 683.
\bibitem{SDdifficulty}
K-I.~Aoki, M.~Bando, T.~Kugo, K.~Hasebe and H.~Nakatani,
Prog.~Theor.~Phys. {\bf 81} (1989) 866.
\bibitem{SDbeyond}
K.~Kondo and H.~Nakatani, Nucl.~Phys.~ {\bf B351} (1991) 236;
Prog.~Theor.~Phys. {\bf 88} (1992) 7373.
K.~Kondo, Int.~J.~Mod.~Phys. {\bf A7} (1992) 7239.
\bibitem{wilsonkogut}
K.~G.~Wilson, I.~G.~Kogut, Phys.~Rep. {\bf 12} (1974) 75.
\bibitem{wegnerhoughton}
F.~Wegner, A.~Houghton, Phys.~Rev. {\bf A8} (1973) 401.
\bibitem{exactRG}
J.~Polchinski, Nucl.~Phys. {\bf B231} (1984) 269.
G.~Keller, C.~Kopper and M.~Salmhofer, Helv.~Phys.~Acta {\bf 65} (1992) 32.
C.~Wetterich, Phys.~Lett. {\bf B301} (1993) 90.
M.~Bonini, M.~D'Attanasio and G.~Marchesini, Nucl.Phys. {\bf B409} (1993) 441.
T.~R.~Morris, Int.~J.~Mod.~Phys, {\bf A9} (1994) 2411.
\bibitem{quarkmeson}
D.~U.~Jungnickel and C.~Wetterich, Phys.~Lett.~{\bf B389} (1996) 600;
Eur.~Phys.~J.~{\bf C2} (1998) 557.
B.~Bergerhoff and C~.Wetterich, Phys.~Rev.~{\bf D57} (1998) 1591.
\bibitem{comoving}
M.~Alford, Phys.~Lett, {\bf B336} (1994) 237.
N.~Tetradis and C.~Wetterich, Nucl.~Phys. {\bf B422} (1994) 541.
T.~R.~Morris, Phys.~Lett. {\bf B329} (1994) 241.
\bibitem{souma}
K-I.~Aoki, K.~Morikawa, W.~Souma, J-I.~Sumi and H.~Terao,
Prog.~Theor.~Phys. {\bf 95} (1996) 409.
\bibitem{fermiliquid}
G.~Benfatto and G.~Gallavotti, J.~Stat.~Phys.~{\bf 59} (1990) 541.
R.~Shankar, Rev. Mod. Phys. {\bf 66} (1993) 129. 
J.~Polchinski, in Proceedings of the 1992 TASI, eds, 
J.~Harvey and J.~Polcjinski (World Scientific, Singapore 1993).
\bibitem{denseqcd}
N.~Evans, S.~D.H.~Hsu and M.~Schwetz, Nucl.~Phys. {\bf B551} (1999)
275; Phys.~Lett. {\bf B449} (1999) 281.
T. Sch\"{a}fer anf F. Wilczek, Phys.~Lett. {\bf B450} (1999) 325.
\bibitem{ccbqed}
K-I.~Aoki, K.~Morikawa, W.~Souma, J.-I.~Sumi and H.~Terao,
Prog.~Theor.~Phys. {\bf 97} (1997) 479.
\bibitem{lpa}
A.~Hazenfratz and P.~Hazenfratz, Nucl.~Phys. {\bf B270} (1986) 269.
T.~R.~Morris, Phys.~Lett, {\bf B334} (1994) 355.
\bibitem{yukawaRG} 
M.~Maggiore, Z.~Phys.~{\bf C41} (1989) 687.
T.E.~Clark, B.~Haeri and S.T.~Love, Nucl.~Phys.~{\bf B402} (1993) 628.
\bibitem{MSTI}
C.~Becchi, {\it On the construction of renormalized quantum field
theory using renormalization group techniques}, in: Elementary
Particles, Field Theory and Statistical Mechanics, eds. M.~Bonini,
G.~Marchesini and E.~Onofri, Parma University, 1993.
U.~Ellwanger, Phys.~Lett. {\bf B335} (1994) 364.
U.~Ellwanger, M.~Hirsch, and A.~Weber, Z.~Phys. {\bf C69} (1996) 687.
M.~Bonini, M.~D'Attanasio, and G.~Marchesini, Nucl.~Phys. {\bf B418}
(1994) 81; {\bf B421} (1994) 429; {\bf B437} (1995) 163;
Phys. Lett. {\bf B346} (1995) 87.
M.~D'Attanasio and T.~R.~Morris, Phys.~Lett. {\bf B378} (1996) 213.
F.~Freire and C.~Wetterich, Phys.~Lett. {\bf B380} (1996) 337.
\bibitem{ksy}
K.~Kondo, S.~Shuto and K.~Yamawaki, Mod.~Phys.~Lett. {\bf A6} (1991)
3385.
K.~Kondo, M.~Tanabashi and K.~Yamawaki, Prog.~Thoer.~Phys. 
{\bf 89} (1993) 1249.
\bibitem{hkkn}
M.~Harada, Y.~Kikukawa, T.~Kugo and H.~Nakano, Prog.~Theor.~Phys. {\bf 92}
(1994) 1161.
N.~V.~Krasnikov, Mod.~Phys.~Lett. {\bf A8} (1993) 797.
\bibitem{paper3}
K-I.~Kubota and H.~Terao, KANAZAWA-99-13.
\bibitem{paper2}
K-I.~Aoki, K.~Morikawa, J.-I.~Sumi, H.~Terao and M.~Tomoyose,
KANAZAWA-99-12, KUCP0140. 
\end{thebibliography}
\end{document}